\documentclass[pra,aps,superscriptaddress,twocolumn]{revtex4-1}

\usepackage{graphicx}
\usepackage{amsmath}
\usepackage{amssymb}
\usepackage{mathbbol}

\begin{document}

\date{January 14, 2015}

\title{Operational Entanglement Families of Symmetric Mixed $N$-Qubit States}

\author{T. Bastin}
\affiliation{Institut de Physique Nucl\'eaire, Atomique et de Spectroscopie, University of Liege, B-4000 Li\`ege, Belgium}

\author{P. Mathonet}
\affiliation{Institut de Math\'ematique, University of Liege, B-4000 Li\`ege, Belgium}

\author{E. Solano}
\affiliation{Department of Physical Chemistry, University of the Basque Country UPV/EHU, Apartado 644, E-48080 Bilbao, Spain}
\affiliation{IKERBASQUE, Basque Foundation for Science, Maria Diaz de Haro 3, E-48013 Bilbao, Spain}

\begin{abstract}
We introduce an operational entanglement classification of symmetric mixed states for an arbitrary number of qubits based on stochastic local operations assisted with classical communication (SLOCC operations). We define families of SLOCC entanglement classes successively embedded into each other, we prove that they are of non-zero measure, and we construct witness operators to distinguish them. Moreover, we discuss how arbitrary symmetric mixed states can be realized in the lab via a one-to-one correspondence between well-defined sets of controllable parameters and the corresponding entanglement families.
\end{abstract}


\maketitle

\section{Introduction}

Entanglement is at the heart of quantum information theory~\cite{Nie00} and its classification is of fundamental interest since all entangled states are not equivalent to each other in terms of their applications~\cite{Horodecki,Guh09}. An important classification scheme in this respect is provided by the classes of states interconvertible between each other through stochastic local operations assisted with classical communication, the so-called SLOCC classification. In this scheme, separable states do always form a unique SLOCC class. For 2- and 3-qubit systems, the number of SLOCC entanglement classes is finite, 2 and 6, respectively, if one includes the class of separable states~\cite{Dur00}. For 3-qubits, 2 of the 6 classes correspond to genuinely entangled states, namely the well-known W and Greenberger-Horne-Zeilinger (GHZ) classes~\cite{Dur00}. For more than three qubits there is an infinite number of SLOCC classes and the introduction of families of entanglement classes~\cite{Ver02} has been instrumental to shed light into the increasing level of complexity~\cite{Lam07}. Recently the concept of entanglement polytopes has provided an insightful geometrical information to discriminate the SLOCC classes~\cite{Wal13}. Ratios of polynomial invariants can also be a very useful tool in this context~\cite{Gou13}. The important subset of symmetric states with respect to permutation of the parties has proven to be easier to tackle and a global classification scheme has been specifically developed for it~\cite{Bas09a}. The proposed scheme fulfills an additional operational criterion where one is able to univocally associate physical knobs in given setups with the defined entanglement families~\cite{Bas09a, Kie10, Lam13}. It has been applied in a variety of further studies~\cite{furtherStudies}.

The case of mixed states is more elaborate and, for three qubits, the notions of compactness and convexity of sets of states proved to be useful~\cite{Aci01}. In this paper, we introduce a classification of mixed symmetric $N$-qubit states into different families of entanglement classes, generalizing the pure state case~\cite{Bas09a}. Our proposal is based on embedded compact convex sets, allowing for the construction of witness operators for each entanglement family and introducing a natural hierarchy between them. In this sense, it offers a full generalization of the three-qubit mixed state classification of Ref.~\cite{Aci01} to the $N$-qubit symmetric case. Here, symmetric mixed states are meant in the sense of Ref.~\cite{Tot09}, i.e., states that can be written as convex sums of projectors onto symmetric pure states. They form a subset of all permutationnally invariant mixed states~\cite{Tot09}. For instance, the mixed state $1/2(|01\rangle\langle01| + |10\rangle\langle10|)$ is permutationnally invariant, but not symmetric since it is an equally weighted convex sum of both the symmetric $(|01\rangle + |10\rangle)/\sqrt{2}$ and antisymmetric $(|01\rangle - |10\rangle)/\sqrt{2}$ states. Therefore, it is not addressed in this study.

The paper is organized as follows. In Sec.~II, the mixed-state entanglement classification is introduced. We then show in Sec.~III how standard projector-based witness operators can be defined for this classification. In Sec.~IV, we explain how our mixed-state classification can be implemented in the lab with one-to-one correspondence between given experimental parameters and the entanglement families. Finally, we draw our conclusions in Sec.~V.

\section{Mixed-state entanglement classification}

Any $N$-qubit symmetric pure state $|\psi_S\rangle$ can always be expressed in terms of $N$ single qubit states $|\epsilon_1\rangle, \ldots, |\epsilon_N\rangle$ in the so-called Majorana representation form~\cite{Maj32, Bas09a}
\begin{equation}
\label{psiS} |\psi_S\rangle = \mathcal{N} \sum_{\pi} |\epsilon_{\pi(1)}, \ldots ,
\epsilon_{\pi(N)}\rangle,
\end{equation}
where the sum runs over all distinct permutations $\pi$ of the qubits and $\mathcal{N}$ is a normalization prefactor. In Eq.~(\ref{psiS}), some $|\epsilon_i\rangle$'s can be identical and are said in this case to have a multiplicity $n_i$ larger than one. The list $\ell$ of multiplicities of each \emph{distinct} single-qubit state $|\epsilon_i\rangle$, sorted by decreasing order, yields the \emph{degeneracy configuration} $\mathcal{D}_{\ell}$ of the symmetric state~$|\psi_S\rangle$ and the number of these distinct states $|\epsilon_i\rangle$ defines the \emph{diversity degree} $d$ of $|\psi_S\rangle$~\cite{Bas09a}. As we have $\sum_i n_i = N$, the degeneracy configuration is nothing else than a partition of the integer $N$ (a list of positive integers that sum to $N$) and the number of distinct degeneracy configurations for a given $N$ is merely given by the number of ways the integer $N$ can be partitioned, the so-called partition function $p(N)$. For instance, a symmetric $N$-qubit state with all $\epsilon_i$ identical is a state with a degeneracy configuration $\mathcal{D}_N$ and a diversity degree $d=1$. A state with all but one $\epsilon_i$ identical has a degeneracy configuration $\mathcal{D}_{N-1,1}$ and a diversity degree $d=2$. If all but two $\epsilon_i$ are the same, the degeneracy configuration of the state is
$\mathcal{D}_{N-2,2}$ ($d=2$) or $\mathcal{D}_{N-2,1,1}$ ($d=3$)
depending on whether the two remaining $\epsilon_i$ are identical or not,
respectively. If all $\epsilon_i$ are distinct, the symmetric state has the degeneracy configuration $\mathcal{D}_{1,\ldots, 1}$ and the highest possible diversity degree $d=N$.

Equation~(\ref{psiS}) offers the advantage of allowing for an easy operational classification of all symmetric pure states, given that the degeneracy configuration $\mathcal{D}_{\ell}$ is an invariant in each SLOCC class~\cite{Bas09a}. This invariant is not necessarily different for the different SLOCC classes~\cite{Bas09a} and a coarse-grained classification of states is defined by the families of SLOCC classes sharing the same degeneracy configuration, the so-called $\mathcal{D}_{\ell}$ families of states. In contrast to the SLOCC classification, the number of such families is always finite for any $N$ and simply given by the partition number $p(N)$~\cite{Bas09a}. The GHZ SLOCC class for symmetric states is always contained in the $\mathcal{D}_{1,\ldots,1}$ family ($d=N$), the Dicke state $|D_N^{(k)}\rangle$ SLOCC classes ($k = 1,\ldots,\lfloor N/2 \rfloor$)~\cite{footnote1} are identified with the $\mathcal{D}_{N-k,k}$ families ($d=2$), while the symmetric separable states define the $\mathcal{D}_{N}$ family ($d=1$)~\cite{Bas09a}. Consequently, we denote these families by GHZ, W$_k$ and S, respectively. In a sense, the $\mathcal{D}_{1,\ldots,1}$ family gathers the most complicated symmetric states characterized with all distinct $\epsilon_i$'s, while $\mathcal{D}_{N}$ the simplest states with all $\epsilon_i$'s identical.

Mixed-state entanglement classification can be obtained generalizing the case of pure states. This was achieved in the three-qubit case~\cite{Aci01} by defining in the real Hilbert space of hermitian operators successive compact and convex sets (classes) of mixed states (density operators) embedded into each other, allowing for the construction of witness operators that are able to distinguish them. In this sense, a prior identification of a hierarchy between all pure three-qubit classes was required. This yielded a scheme of successive closed sets of mixed states embedded into each other, namely $\mathrm{S} \subset \mathrm{B} \subset \mathrm{W} \subset \mathrm{GHZ}$, $\mathrm{B}$ denoting the set of biseparable states. We fully generalize here this approach to the symmetric subspace of the $N$-qubit case. To this aim, we identify with the same approach as in Ref.~\cite{Aci01} the unique hierarchy between all aforementioned $\mathcal{D}$ families of symmetric pure states that can lead to a comprehensive classification of mixed states.
We first exemplify this for $N = 4$. In this case, the symmetric pure states (\ref{psiS}) are fully determined by the knowledge of four $\epsilon_i$'s, yielding the five entanglement families $\mathcal{D}_{1,1,1,1}$, $\mathcal{D}_{2,1,1}$, $\mathcal{D}_{2,2}$, $\mathcal{D}_{3,1}$, and $\mathcal{D}_{4}$~\cite{Bas09a}. The first family gathers symmetric states characterized with four distinct $\epsilon_i$'s and is spanned by considering all quadruplets $(\epsilon_1,\epsilon_2,\epsilon_3,\epsilon_4)$ fulfilling the condition of distinctness. This condition makes the set open since each state of this set is necessarily at a minimal nonzero distance of any states out of the set which have two or more identical $\epsilon_i$'s~\cite{footnote2}. These latter states $\notin \mathcal{D}_{1,1,1,1}$ belong to any of the four remaining families and are in contrast at an arbitrarily small distance of $\mathcal{D}_{1,1,1,1}$ states. They all lie consequently at the boundary of the $\mathcal{D}_{1,1,1,1}$ set~\cite{footnote2}. Along the same lines, one observes how the other successive families fit into each other: $\mathcal{D}_{2,2}$, $\mathcal{D}_{3,1}$, and $\mathcal{D}_{4}$ lie at the boundary of the $\mathcal{D}_{2,1,1}$ family, and $\mathcal{D}_{4}$ is at the boundary of $\mathcal{D}_{2,2}$ and $\mathcal{D}_{3,1}$. Note that neither $\mathcal{D}_{2,2}$ is at the boundary of $\mathcal{D}_{3,1}$ nor the converse. The family $\mathcal{D}_{1,1,1,1}$ is the only open set and $\mathcal{D}_{4}$ the only closed one.

This highlights the natural hierarchy of the entanglement families when they are ordered according to their boundary imbrication, a family $\mathcal{D}'$ being said to \emph{descend} from $\mathcal{D}$ ($\mathcal{D} \rightarrow \mathcal{D}'$) if $\mathcal{D}'$ lies at the boundary of $\mathcal{D}$. In the general $N$ case, the $\mathcal{D}_{1,\ldots,1}$ family is always the primary set at the boundary of which all others reside and $\mathcal{D}_{N}$ is located in the other extreme at the boundary of all other families. The only open set is $\mathcal{D}_{1,\ldots,1}$ and $\mathcal{D}_N$ is the only closed one. All descendants of a family $\mathcal{D}$ form its complete boundary and the union of \emph{all} these descendants with $\mathcal{D}$ thus yields the closure $\overline{\mathcal{D}}$ of $\mathcal{D}$~\cite{footnote2}. We have for instance in the $N=4$ case $\overline{\mathcal{D}}_{2,1,1} = \mathcal{D}_{2,1,1} \cup \mathcal{D}_{3,1} \cup \mathcal{D}_{2,2} \cup \mathcal{D}_{4}$. As a consequence, any continuous entanglement measure that would vanish for all states of a given family $\mathcal{D}$ would also vanish for all states in all descending families since they belong to the closure $\overline{\mathcal{D}}$.

\begin{figure}
\begin{center}
\includegraphics[width = 6 cm, bb=70 350 420 670, clip=true]{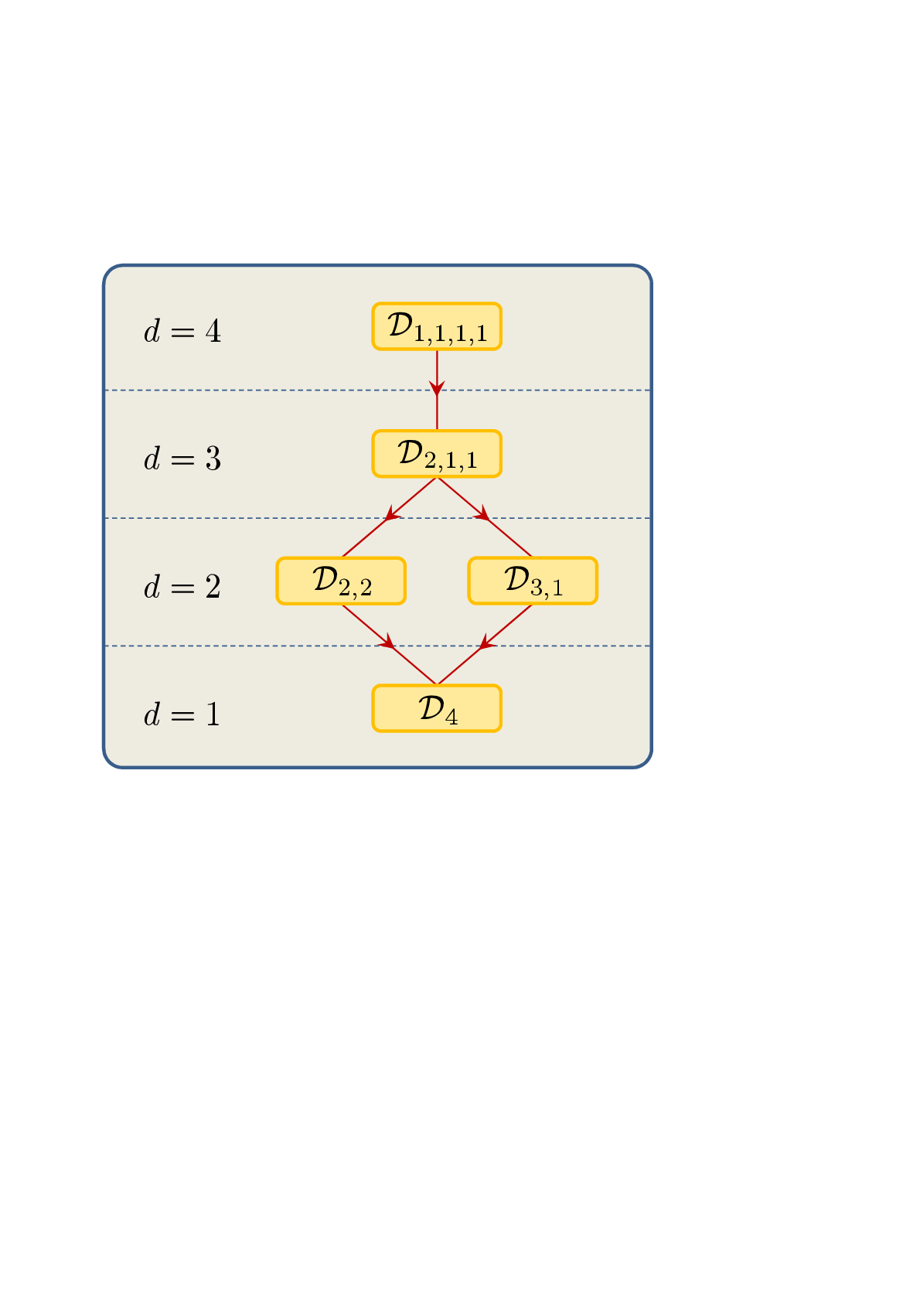}
  \caption{(Color online) Pure-state entanglement family graph for $N = 4$.}
	\label{FigN4e}
\end{center}
\end{figure}

The family hierarchy can be advantageously illustrated using an \emph{entanglement family graph}, as shown in Fig.~\ref{FigN4e} for $N=4$. In these graphs, any downward arrow or path of downward arrows materializes a descending relation between two families. In the general $N$ case, $\mathcal{D}_{1,\ldots,1}$ is always the highest level family with respect to the defined hierarchy and $\mathcal{D}_{N}$ the lowest level one with no descendants. Lower level families have necessarily lower diversity degrees. Any family with a given diversity degree $d$ never descends from another one with same $d$, while it always descends from at least a family with diversity degree $d+1$ (for $d < N$). Consequently, the diversity degree is a good hierarchy marker and it makes sense to display the entanglement family graphs by layers of families of up-to-bottom diversity degree. The number of layers is exactly $N$ and the layers $d=N$, $d=N-1$, and $d=1$ never contain more than one entanglement family, namely $\mathcal{D}_{1,\ldots,1}$, $\mathcal{D}_{2,1,\ldots,1}$, and $\mathcal{D}_{N}$, respectively.

Symmetric $N$-qubit mixed states are here considered as states that can be written as convex sums of projectors onto symmetric pure states~\cite{Tot09}. These states can be classified generalizing the $\mathcal{D}$ family classification of the latter and taking into account the hierarchy of these families, similarly to the approach followed in Ref.~\cite{Aci01} for the specific 3-qubit case. To this aim, we define the $\mathcal{C}_{\ell}$ families in the space of density operators as being the sets of symmetric mixed states that can be written as convex sums of projectors onto symmetric pure states of the respective $\mathcal{D}_{\ell}$ families \emph{and any of their descendants}, i.e., onto symmetric pure states of the respective closures $\overline{\mathcal{D}}_{\ell}$. In this manner, the formed sets of mixed states are compact and convex~\cite{footnote3}, allowing for the construction of witness operators~\cite{Guh09} detecting states outside the sets according to the Hahn-Banach separation theorem. Clearly we have furthemore
\begin{equation}
\mathcal{C}_{\ell'} \subset \mathcal{C}_{\ell}
\end{equation}
for all $\ell,\ell'$ such that $\mathcal{D}_{\ell'}$ descends from $\mathcal{D}_{\ell}$. In this case $\mathcal{C}_{\ell'}$ is said to be a subfamily of $\mathcal{C}_{\ell}$. This results in an onion-like layer structure of the different $\mathcal{C}$ families of mixed states with a complexity growing with that of the entanglement family graph. The physical interpretation of the classification is as follows: a mixed state $\hat{\rho}$ that belongs to a $\mathcal{C}_{\ell}$ family and to none of the subfamilies is a mixed state whose simplest decompositions into projectors onto symmetric pure states require at least a pure symmetric state of the $\mathcal{D}_{\ell}$ family. For any $N$, $\mathcal{C}_N$ is the only family that has no subfamilies and it gathers consequently mixed states that can be written as convex sums of only $\mathcal{D}_N$ pure states, that is of only separable states. This family is nothing else than the usual set of separable mixed states restricted to the symmetric subspace.

Let us make explicit our mixed-state classification for $N=2, 3$ and $4$, while the case of higher $N$'s extrapolates straightforwardly. For up to three qubits, it fits with well-known results of mixed-state entanglement classification, namely with the non-substructure of two-qubit entanglement and with the Ac\'in {\it et al.}~\cite{Aci01} classification for mixed three-qubit states. Beyond these cases, it yields a general entanglement classification of higher number of qubits when focusing on mixed symmetric states.
For $N = 2$, there are only two pure-state families, $\mathcal{D}_{1,1}$ ($d = 2$) and $\mathcal{D}_2$ ($d=1$)~\cite{Bas09a}, gathering symmetric entangled and separable states, respectively, with hierarchy $\mathcal{D}_{1,1} \rightarrow \mathcal{D}_{2}$. Accordingly, for mixed states, the $\mathcal{C}_{2}$ family contains all separable symmetric mixed states, while $\mathcal{C}_{1,1}$ contains all mixed states that can be expressed as a convex sum of projectors onto $\mathcal{D}_2$ and $\mathcal{D}_{1,1}$ states, that is onto any symmetric state. The set difference $\mathcal{C}_{1,1} \setminus \mathcal{C}_2$ corresponds merely to the set of entangled symmetric mixed states. We have $\mathcal{C}_2 \subset \mathcal{C}_{1,1}$ and this set structure reflects merely the usual simple structure of 2-qubit mixed state entanglement according to which the separable state set is surrounded by the entangled state set, the latter one showing no substructure.

For $N = 3$, we have the three pure-state entanglement families, $\mathcal{D}_{1,1,1}$ ($d = 3$), $\mathcal{D}_{2,1}$ ($d=2$), and $\mathcal{D}_{3}$ ($d=1$), identifying the GHZ, W, and separable SLOCC classes for symmetric states, respectively~\cite{Bas09a}. The hierarchy of the 3 families forms the chain $\mathcal{D}_{1,1,1} \rightarrow \mathcal{D}_{2,1} \rightarrow \mathcal{D}_{3}$. Accordingly, the $\mathcal{C}_{3}$ mixed-state family contains all states that can be expressed as a convex sum of projectors onto $\mathcal{D}_3$ states only; the $\mathcal{C}_{2,1}$ mixed-state family contains all states that can be expressed as a convex sum of projectors onto $\mathcal{D}_3$ and $\mathcal{D}_{2,1}$ states; and the $\mathcal{C}_{1,1,1}$ mixed-state family contains all states that can be expressed as a convex sum of projectors onto $\mathcal{D}_3$, $\mathcal{D}_{2,1}$ and $\mathcal{D}_{1,1,1}$ states, that is onto any symmetric state.  The $\mathcal{C}_{1,1,1}$, $\mathcal{C}_{2,1}$, and $\mathcal{C}_{3}$ families for mixed states are nothing else than the GHZ, W, and separable S classes of mixed states identified in Ref.~\cite{Aci01}, respectively, disregarding the biseparable sets nonexistent in the symmetric subspace. The resulting structure $\mathcal{C}_{3} \subset \mathcal{C}_{2,1} \subset \mathcal{C}_{1,1,1}$ is just the translation of the relation $\mathrm{S} \subset \mathrm{W} \subset \mathrm{GHZ}$~\cite{Aci01} for symmetric states.

\begin{figure}
\begin{center}
\includegraphics[width = 6cm, bb=115 275 510 520, clip=true]{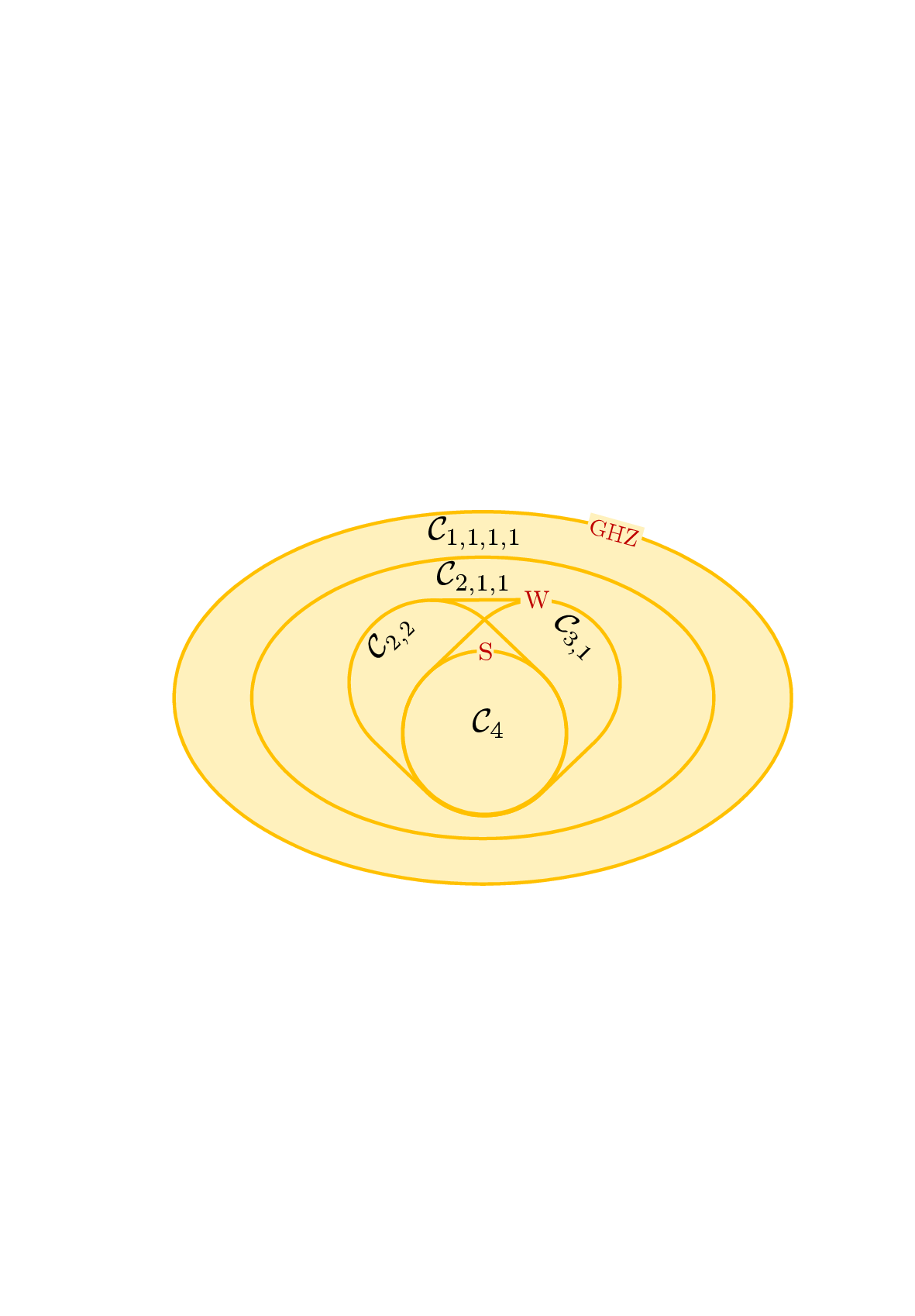}
  \caption{(Color online) Symmetric mixed-state family structure for $N = 4$ (see text).}
	\label{FigN4}
\end{center}
\end{figure}

For $N = 4$, we have the five entanglement pure-state $\mathcal{D}$ families as shown in Fig.~\ref{FigN4e}, along with their hierarchy which starts to be more involved than the simple direct structure of the cases $N=2$ and $3$. Consequently, for mixed states, the $\mathcal{C}_4$ family contains all states that can be expressed as a convex sum of projectors onto $\mathcal{D}_4$ states; the $\mathcal{C}_{3,1}$ family contains all states that can be expressed as a convex sum of projectors onto $\mathcal{D}_4$ and $\mathcal{D}_{3,1}$ states; $\mathcal{C}_{2,2}$ contains all states that can be expressed as a convex sum of projectors onto $\mathcal{D}_4$ and $\mathcal{D}_{2,2}$ states; $\mathcal{C}_{2,1,1}$ contains all states that can be expressed as a convex sum of projectors onto $\mathcal{D}_4$, $\mathcal{D}_{3,1}$, $\mathcal{D}_{2,2}$, and $\mathcal{D}_{2,1,1}$ states; and, finally, the $\mathcal{C}_{1,1,1,1}$ mixed-state family contains all states that can be expressed as a convex sum of projectors onto any symmetric state. It follows the family set structure as shown in Fig.~\ref{FigN4}. The convex hull of $\mathcal{C}_{3,1}$ and $\mathcal{C}_{2,2}$ collects mixed states that can be written as convex sums of $\mathcal{D}_4$, $\mathcal{D}_{3,1}$, and $\mathcal{D}_{2,2}$ states, that is on S and W-kind states. This is why it is denoted by W in Fig.~\ref{FigN4}. It is also \emph{a priori} not excluded that a state belong at the same time to $\mathcal{C}_{3,1}$ and to $\mathcal{C}_{2,2}$ while not being separable, i.e., that a state be in the intersection of $\mathcal{C}_{3,1}$ and $\mathcal{C}_{2,2}$ minus $\mathcal{C}_4$. We would have in this case a non-separable state that could be expressed as a convex sum of projectors onto either $\mathcal{D}_{3,1}$ and $\mathcal{D}_{4}$ states, or $\mathcal{D}_{2,2}$ and $\mathcal{D}_{4}$ states. A similar behavior is for instance observed for nonsymmetric three-qubit states where mixed states can be found biseparable with respect to any bipartition without being fully separable~\cite{Aci01,Guh09}.

For pure states, all $\mathcal{D}$ entanglement families form zero-measure sets with the exception of $\mathcal{D}_{1,\ldots,1}$ at the boundary of which all other families reside. For mixed states, the situation is totally different and all $\mathcal{C}$ families are of non-zero measure. This can be proven as follows. The symmetric $N$-qubit mixed state density operators belong to the $(N+1)^2$-dimensional real Hilbert space of symmetric hermitian operators. A natural basis in this space is given by the set of operators
\begin{equation}
\label{basis}
\{ \hat{\sigma}_\lambda \equiv \hat{\sigma}_{N,k}, \hat{\sigma}_{N,k,j}^{(r)}, \hat{\sigma}_{N,k,j}^{(i)}, \, j,k=0,\ldots,N, j>k\},
\end{equation}
with
$\hat{\sigma}_{N,k} = |D_N^{(k)}\rangle\langle D_N^{(k)}|$, $\hat{\sigma}_{N,k,j}^{(r)} = |D_N^{(k)}\rangle\langle D_N^{(j)}| + |D_N^{(j)}\rangle\langle D_N^{(k)}|$, and $\hat{\sigma}_{N,k,j}^{(i)} = i(|D_N^{(k)}\rangle\langle D_N^{(j)}| - |D_N^{(j)}\rangle\langle D_N^{(k)}|)$.
In this basis, any pure symmetric separable state $|\epsilon\rangle^{\otimes N} \langle \epsilon|$ with $|\epsilon\rangle = \cos(\theta/2)|0\rangle + \sin(\theta/2)e^{i\phi}|1\rangle$ reads
$|\epsilon\rangle^{\otimes N} \langle \epsilon| = \sum_{\lambda} f_\lambda(\theta,\phi) \hat{\sigma}_\lambda$ with $f_{N,k,j}^{(r)}(\theta,\phi)$ and $f_{N,k,j}^{(i)}(\theta,\phi)$ the real and imaginary part of $(C_N^k C_N^j)^{1/2} \cos(\theta/2)^{2N-(k+j)} \sin(\theta/2)^{k+j} e^{i(k-j)\phi}$, respectively, and $f_{N,k}(\theta,\phi) = f_{N,k,k}^{(r)}(\theta,\phi)$. The $(N+1)^2$ functions $f_\lambda(\theta,\phi)$ are easily seen to be linearly independent and this implies that the symmetric separable states span the Hilbert space of symmetric hermitian operators. In particular, a (non-orthonormal) basis of $(N+1)^2$ such separable states can be found and their affine hull (set of all their linear combinations with coefficients, positive or negative, adding up to 1) yields the subset of all trace one operators containing in particular all symmetric mixed states. Since the convex hull of a set is of non-zero measure inside its affine hull, the convex hull of the $(N+1)^2$ separable basis states is a non-zero measure set of symmetric separable states inside the whole symmetric mixed state space. The conclusion follows that the $\mathcal{C}_N$ family is of non-zero measure. All other families being convex and compact, embedded into each others and containing $\mathcal{C}_N$, each successive family adds a non-zero measure set of states with respect to all their descendants, as a direct consequence of the Hahn-Banach separation theorem.

\section{Witness operators}

Having established the structure of the set of symmetric mixed $N$-qubit states, we now show how witness operators can be used to distinguish these different families of multipartite entanglement. Witness operators are useful tools that allow one to detect states lying outside compact convex sets~\cite{Guh09}. Thanks to this property of the defined $\mathcal{C}$ families, witnesses can always be built to detect mixed states lying out of them and thus belonging to their complements, that is to say to any higher or similar level family. For a given $\mathcal{C}$ family, such witnesses are observables whose expectation values are positive for any state belonging to $\mathcal{C}$ and strictly negative for at least a state outside $\mathcal{C}$, that is inside the complement set $\mathcal{C}^c$~\cite{Guh09}. We call these observables $\mathcal{C}^c$-witnesses and denote them by $\mathcal{W}_{\mathcal{C}^c}$. Any symmetric state $\hat{\rho}$ fulfilling $\mathrm{Tr}(\mathcal{W}_{\mathcal{C}^c} \hat{\rho}) < 0$ is said to be detected by the witness and is guaranteed to belong to $\mathcal{C}^c$ and therefore not to be a $\mathcal{C}$-state. As witnesses cannot detect states lying inside compact convex sets, $\mathcal{C}$-witness operators do not exist. Since $(\mathcal{C}_N)^c$-witnesses detect non-separable states they are just entanglement witnesses in the symmetric subspace~\cite{Tot10}.

Projector-based operators provide a convenient way for building $\mathcal{C}^c$-witnesses. Any observable of the form
\begin{equation}
    \mathcal{W} = \alpha_{\mathcal{D}_l} \mathbb{1} - |\psi\rangle \langle \psi|,
\end{equation}
with $|\psi\rangle$ a state out of $\mathcal{D}_l$ and
\begin{equation}
    \alpha_{\mathcal{D}_l} = \mathrm{max}_{|\phi\rangle \in \mathcal{D}_l} |\langle \phi | \psi \rangle |^2
\end{equation}
is a $(\mathcal{C}_l)^c$-witness. Let us exemplify different family witnesses for $N=4$. First we consider witness operators built using projectors onto the GHZ state. In this state, we do have $\alpha_{\mathcal{D}_4} = 1/2$~\cite{Wei03}, $\alpha_{\mathcal{D}_{3,1}} = 1/2$, $\alpha_{\mathcal{D}_{2,2}} = 3/4$ and $\alpha_{\mathcal{D}_{2,1,1}} = 7/8$. Since $\alpha_{\mathcal{D}_{3,1}} = \alpha_{\mathcal{D}_4}$, the GHZ state does not provide distinct $\mathcal{C}^c_{3,1}$- and $\mathcal{C}^c_{4}$-witnesses. A better option with that concern is to consider witnesses based on projectors onto the ``tetrahedron state''
$|T_4\rangle = 1/\sqrt{3} |D_4^{(0)}\rangle + \sqrt{2/3} |D_4^{(3)}\rangle$, so called because the Bloch sphere representation of its four $|\epsilon_i\rangle$ states [see Eq.~(\ref{psiS})] do form a regular tetrahedron~\cite{Mar10}. For $|T_4\rangle$, the maximal overlaps with the different family states read $\alpha_{\mathcal{D}_4} = 1/3$~\cite{Mar10}, $\alpha_{\mathcal{D}_{3,1}} = 2/3$, $\alpha_{\mathcal{D}_{2,2}} = 1/2$ and $\alpha_{\mathcal{D}_{2,1,1}} = 3/4$, allowing ones to witness distinctly all four-qubit family complement sets.

\section{Operational implementation}

The proposed classification of mixed entangled states in the symmetric subspace can be implemented in the lab with a one-to-one correspondance between given experimental parameters and the entanglement families. In the experimental setups proposed in Refs.~\cite{Bas09b}, $N$-qubit symmetric pure states are produced in atomic and photonic systems by initial preparation or projective measurements of $N$ individual photon polarization states using adequate elliptical polarizers. There, a final atomic or photonic state of the form~(\ref{psiS}) is produced, where each qubit state $|\epsilon_i\rangle = \alpha_i |0\rangle + \beta_i |1\rangle$ is directly determined by the polarization orientation $\boldsymbol{\epsilon}_i = \alpha_i \boldsymbol{\sigma}_- + \beta_i \boldsymbol{\sigma}_+$ of the $i$th polarization filter. A one-to-one correspondance between the degeneracy configuration of the polarizer orientations (list of numbers of identical polarizers) and the $\mathcal{D}$ family of the pure state produced in the atomic or photonic $N$-qubit system is obtained. This correspondance is generalized to the symmetric mixed states by considering the polarizer orientations to be unknown within the perimeter imposed by a given degeneracy configuration. For instance, in a four-qubit set-up, three identical polarizer orientations distinct from a last one while ignoring their exact orientations produces a $\mathcal{C}_{3,1}$ symmetric mixed state, whose decomposition onto projectors of symmetric pure states is known from the probability distribution of the allowed polarizer orientations, each orientation set determining a well-defined pure state~\cite{Bas09b}. It is noteworthy to mention that with this approach all symmetric mixed multiqubit states can be produced.

\section{Conclusion}

We have introduced an operational entanglement classification of symmetric mixed states for an arbitrary number of qubits based on families of SLOCC entanglement classes successively embedded into each other. We have proven that all these families are of non-zero measure and that witness operators can be defined to distinguish them. Finally, we have shown how arbitrary symmetric mixed states can be realized in the lab via a one-to-one correspondence between well-defined sets of controllable parameters and the proposed entanglement families.

\acknowledgments
T.B. acknowledges the financial support from the Belgian FRS-F.N.R.S. through IISN Grant 4.4512.08.
E.S. acknowledges support from the Spanish MINECO FIS2012-36673-C03-02; UPV/EHU UFI 11/55; Basque Government IT472-10; CCQED, PROMISCE, SCALEQIT EU projects. It is a pleasure to thank O. G\"uhne and G. T\'oth for helpful discussions.


\begin{thebibliography}{99}
\bibitem{footnote3}
In finite dimension, a compact set of states is a closed and bounded set. The closeness condition is actually sufficient since $\mathrm{Tr}(\hat{\rho}^2) \leqslant 1$ for all states $\hat{\rho}$. A set $\mathcal{A}$ of mixed states is convex iff $p \hat{\rho} + (1-p) \hat{\rho}' \in \mathcal{A}$ for all $\hat{\rho}, \hat{\rho}' \in \mathcal{A}$ and $p \in [0,1]$.
\end{thebibliography}
\end{document}